\documentclass[twocolumn,aps,superscriptaddress,longbibliography,amsmath,amssymb,reprint]{revtex4-2}
\usepackage{amsmath}  % Include the AMS math package

\usepackage{graphicx}% Include figure files
\usepackage{dcolumn}% Align table columns on decimal point
\usepackage{bm}% bold math
\usepackage{hyperref}
\usepackage{siunitx}
\usepackage{color}
\usepackage[normalem]{ulem}

\begin{document}

\preprint{APS/123-QED}

\title{Atomistic study of finite temperature properties in ferroelectric BiAlO$_3$}

\author{Indranil Mal}
%\email{indranil.mal@gmail.com}
\affiliation{Department of Physics, Indian Institute of Technology,
Hauz Khas, New Delhi 110016, India}
\author{Chandan Kumar Vishwakarma}
\affiliation{Department of Physics, Indian Institute of Technology,
Hauz Khas, New Delhi 110016, India}
\author{Mohd Zeeshan}
\affiliation{Department of Physics, Indian Institute of Technology,
             Hauz Khas, New Delhi 110016, India}
\author{I. Ponomareva}
\affiliation{Department of Physics, University of South Florida,
            Tampa, Florida 33620, USA}
\author{B. K. Mani}
\email{bkmani@physics.iitd.ac.in}
\affiliation{Department of Physics, Indian Institute of Technology,
Hauz Khas, New Delhi 110016, India}

%\date{\today}

\begin{abstract}

The lead-free perovskite ferroelectrics captivate researchers with their 
unique functional properties leading to important technological applications.
In the search for a new lead-free perovskite of technological importance, we 
develop a first-principles based atomistic model to accurately predict the 
properties of BiAlO$_3$ in experimentally relevant conditions. Consistent 
with the experimental observations, our simulations predict a rhombohedral 
ferroelectric ($R3c$) ground state for BiAlO$_3$ facilitated by a structural 
phase transition from paraelectric (cubic, $Pm\Bar{3}m$) phase. The
room-temperature spontaneous polarization and Curie temperature are obtained
to be 81 $\mu$C/cm$^2$ (along [111] direction) and 1160 K, respectively. 
Our simulations reveal strong coupling between ferroelectric and 
antiferrodistortive modes for a broad spectrum of temperature and 
electric field. We find that hydrostatic pressure suppresses both spontaneous 
polarization  and Curie temperature, while both uniaxial and biaxial 
stresses induce  multiple phase transitions in BiAlO$_3$. 

\end{abstract}

\maketitle

\section{Introduction}

Ferroelectric are in the focus of  attention owing to their 
extensive use in numerous technologically important applications such as
non-volatile memory \cite{liao-09, Memory_2024}, piezoelectric \cite{piezo_2000, piezo_2017} 
and electromechanical devices \cite{chemical_2011, xu-98}, 
pyroelectric sensors \cite{pyro_1998}, electro-caloric cooling \cite{scott-11, mischenko-06} 
and many others \cite{Zhang_2016,Zhao_2016,Bhatti_2016}. 
Within this class, lead-based perovskites and their solid solutions have established 
themselves as outstanding functional materials for these 
applications \cite{Yadav_2017,Jaouen_2007,Gotthard_1998,mani_atomistic_2013,piezo_2000}. 
However, due to the environmental and health hazards arising from lead toxicity, 
replacing lead-based ferroelectric materials with lead-free alternatives 
is both necessary and desirable, driving extensive efforts toward the 
discovery of new, sustainable ferroelectric compounds \cite{JUN-KI_2007,Eitel_2002,Zvezdin_1994,zylberberg-07,li_novel_2019}.

Recently, bismuth-containing perovskites 
emerged  as promising  alternatives to the lead-based ferroelectrics. 
One representative is BiAlO$_3$. It has been synthesized using high-pressure high-temperature 
techniques \cite{belik-06, zylberberg-07, mangalam-08}.
The X-ray diffraction study \cite{belik-06} has reported a $R3c$ ground 
state structure with lattice parameters $a = 5.38$ \AA\; 
and $c = 13.39$ \AA\;. 
Ferroelectricity in BiAlO$_3$ has been established on the basis of temperature-dependent 
dielectric constant and electric  hysteresis measurements \cite{zylberberg-07,mangalam-08}.  
The experimental Curie temperature is $T_{\rm C}>$ 
793 K, while  room-temperature spontaneous polarization  ranges from $P_{\rm S} = $9.5 
to  11.5 $\mu$C/cm$^2$ \cite{zylberberg-07,mangalam-08}. The room temperature  saturation polarization is $P_{\rm Sat}=$ 
16 $\mu$C/cm$^2$ \cite{mangalam-08}). Another experimental work \cite{son-08} has 
reported ferroelectricity in an epitaxial BiAlO$_3$ thin film 
with a $P_{\rm S}$ of 29 $\mu$C/cm$^2$. It is presently believed that ferroelectric 
instability in BiAlO$_3$ is driven by   stereochemically active 6s$^2$ lone 
pair of electrons in Bi$^{3+}$, similar to Pb$^{2+}$ in prototype 
ferroelectric PbTiO$_3$ \cite{baettig-05}. 
Consistent with experimental observations, first-principles density 
functional theory (DFT) calculations have also predicted BiAlO$_3$ to be promising 
ferroelectric associated with large distortions from the cubic structure driven by 
stereochemical activity of Bi lone pair electrons \cite{korus_22, kaczkowski-16,kang_22, baettig-05}.  
However, as with most DFT calculations, the evaluated properties are restricted to 
zero Kelvin and therefore do not capture the material behavior under experimentally relevant conditions. Consequently, 
there exists a large variation in the polarization values reported 
from DFT calculations and experiments.
While DFT ground state calculations report polarizations in the range 
of 72.3 - 82.5 $\mu$C/cm$^2$ \cite{korus_22, kaczkowski-16, kang_22, baettig-05}, 
experiments show even larger variation 
from 9 - 29 $\mu$C/cm$^2$ \cite{zylberberg-07, li_novel_2019, mangalam-08, son-08}. 
To the best of our knowledge, presently there is no computational approach that 
allows to extend the reach of DFT simulations to finite-temperature in BiAlO$_3$. 
Consequently, there is a lack of insight into  finite-temperature properties of BiAlO$_3$. 
For example, the temperature evolution of spontaneous polarization, strain, coercive 
field, hysteresis loops are  presently unknown. There is a lack of atomistic 
insight into the role of coupling between ferroelectric instability and antiferrodistortive 
oxygen octahedra tilts. In this study we aim to fill these gap by (i) developing 
parameterization for first-principles-based finite-temperature simulations of 
BiAlO$_3$; (ii) applying such simulations to predict finite-temperature properties 
of BiAlO$_3$ such as the nature of phase transition 
and associated order parameters, temperature evolution of coercive field, strain 
and hysteresis, etc.; (iii) extending predictions to the regime where BiAlO$_3$  is 
subjected to external mechanical load, such as hydrostatic pressure and uniaxial and biaxial  stresses.

\section{Methodology}

We begin by developing a parameterization for a first-principles-based effective 
Hamiltonian \cite{zhong-95,vanderbilt-98,mani_pzo_2015} for BiAlO$_3$, which allows to extend 
the reach of DFT simulations to finite temperatures.    To identify the degrees of freedom 
contributing to the effective Hamiltonian, we first examined the structural 
instabilities present in the system with the help of phonon dispersions. 
For this, phonons were computed for cubic phase within the framework of density 
functional perturbative theory (DFPT) as implemented in the VASP 
package \cite{kresse_efficiency_1996,kresse_efficient_1996}. 
For this, we used the projector-augmented-wave (PAW) \cite{blochl_projector_1994} 
method to represent the atomic cores and the local-density-approximation 
(LDA) \cite{ceperley_ground_1980,perdew_self-interaction_1981} as the 
exchange-correlation functional. A dense k-mesh of $11\times11\times11$ and 
an energy cutoff of 600 eV for plane wave basis were used in 
the calculations. All self-consistent-field calculations were allowed to
converge till the energy difference was less than  $10^{-6}$ eV. For structural 
relaxation  we required forces to be less than  $10^{-4}$ eV/\AA\ in magnitude.

%%%%%%%%%%%%%%%%%%%% FIGURE 1 
\begin{figure*}
\begin{center}
  \includegraphics[width = 0.8\textwidth]{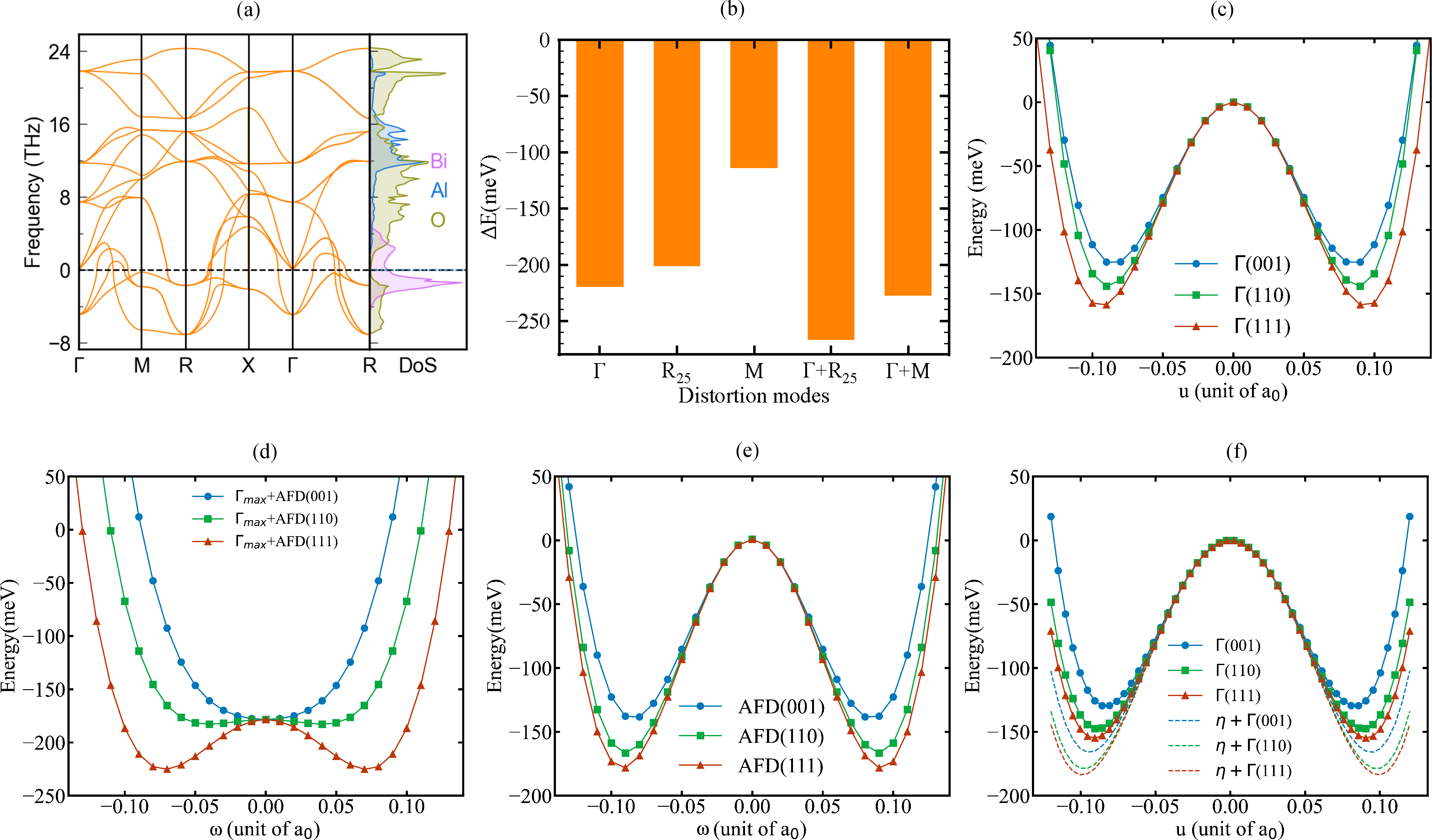}
   \caption{(a) The phonon dispersion and atom projected phonon density of states 
	for BiAlO$_3$. (b) Energy (with respect to the cubic phase) associated 
	with different individual or pair distortions.  (c), (d) and (e) 
	The energy landscapes for $\Gamma$, $R_{25}$ and $\Gamma + R_{25}$ 
	distortions, respectively, in the absence of strain.  (f) 
	The energy landscape for $\Gamma $ point distortion in the presence of strain.}
  \label{fig_ph}
\end{center}
\end{figure*}

The phonon dispersion, along with the partial phonon density of states (pDOS) 
of the constituent atoms, for $Pm\Bar{3}m$ cubic phase of BiAlO$_3$ is shown
in Fig.~\ref{fig_ph}(a). The imaginary frequency phonons in the Brillouin 
zone imply the structural instabilities present in the material. Similar 
features of phonon dispersion were also observed in the previous 
calculation \cite{baettig-05}. 
$\Gamma$ mode corresponds to the opposite displacement of Bi 
atom to Al and O, and is responsible for inducing an electric dipole moment 
in the system. The R and M point instabilities correspond to the tilt 
of oxygen octahedra due to O distortions. 
Panel (b) shows the energy (with respect to the cubic phase) associated with 
different individual and pair distortions. The energy is obtained using 
self-consistent-field calculations using distortions for different unstable phonons. 
The largest gain in energy (219  meV) is associated with  $\Gamma$-point distortion. 
The optimized structure with $R_{25}$ distortions lies  $\sim$ 201 meV 
below the cubic phase and corresponds to a rhombohedral structure with 
$R\bar3c$ symmetry. Similarly, the $M$ point instability is observed to 
be of $P4/mbm$ symmetry and stays   $\sim$ 113 meV below the cubic phase. 
It should, however, be noted that the global minimum structure is achieved 
when $R_{25}$ and $\Gamma$ modes are frozen together and exists $\sim$ 266 meV 
below the undistorted cubic structure. This corresponds to a rhombohedral
structure with $R3c$ space group. 
Our computed energies and corresponding 
structures for different distortions are in good agreement with the previous 
DFT calculations \cite{baettig-05, korus_22}.
Panels (c), (d), and (e)  show the energy landscapes for $\Gamma$, 
$R_{25}$, and $\Gamma+R_{25}$ modes in (001), (110) and (111) 
directions without strain degree of freedom. 
We note that both $\Gamma$ and $R_{25}$ instabilities are associated with 
double well energy landscapes. The most energetically favorable phase is with 
polar direction along $<111>$ and octahedra tilts along the polar axis. 
When both $\Gamma$ and $R_{25}$ instabilities are present the phase with the 
R phase gains in stability a bit, while T phase is destabilized and O 
phase is nearly destabilized. Panel (f) 
shows the energy landscape for the $\Gamma$ modes in the presence 
of strain modes. We find that strain  causes further stabilization of all the phases.
Based on above analysis, we include the dominant distortions $\Gamma$ 
and $R_{25}$ along with the strain degree of freedom in the construction 
of the effective Hamiltonian. The total energy of BiAlO$_3$ can, therefore, be 
expressed in terms of ferroelectric local mode ($u$), antiferrodistortive 
(AFD) mode ($\omega$) (which corresponds to $R_{25}$ distortions) and 
strain ($\eta$) \cite{zhong-95,vanderbilt-98}, as 
\begin{equation}
\begin{aligned} 
E^{\text {tot }}= & E^{\mathrm{FE}}\left(\left\{\mathbf{u}_i\right\}\right)
+ E^{\mathrm{AFD}}\left(\left\{\boldsymbol{\omega}_i\right\}\right) 
+ E^{\text{elas}}\left(\left\{\eta_i\right\}\right) \\
& + E^{\mathrm{FE}-\text { elas }}\left(\left\{\mathbf{u}_i, \eta_i\right\}\right) 
+ E^{\mathrm{AFD}-\text { elas }}\left(\left\{\boldsymbol{\omega}_i, \eta_i\right\}\right) \\ 
&	+ E^{\mathrm{FE}-\mathrm{AFD}}\left(\left\{\mathbf{u}_i, \boldsymbol{\omega}_i\right\}\right).
\end{aligned} 
\label{h_eff}
\end{equation}
Here, $E^{\mathrm{FE}}$ is the energy associated with ferroelectric mode of 
the system and subsumes the energy contributions associated with the local mode 
self-energy, short-range interaction between nieghbouring dipoles and the 
long-range dipolar interactions \cite{zhong-95, vanderbilt-98}. The second term, 
$E^{\mathrm{AFD}}$, represents the energy associated with AFD local mode and 
includes the energy contributions from interactions similar to FE mode 
except the dipolar interaction, as it does not contribute to AFD mode. 
The third term, $E^{\text{elas}}$, is the elastic energy and 
arises due to the deformation of the unit cell. The last three terms represent 
the energy contributions due to the coupling between the FE mode 
and strain, the AFD mode and strain, and the FE and AFD modes, 
respectively \cite{zhong-95, vanderbilt-98}. 
The parameters in Eq.(\ref{h_eff}) are derived from the first-principles 
density functional theory based calculations and are provided in 
Table \ref{tab_param}.

Next, these parameters are used in the framework of single-flip Metropolis Monte Carlo (MC) 
simulations to investigate the properties of BiAlO$_3$ at finite 
temperatures. 
%In this approach, each trial involves updating a single variable 
%(local mode in the present case), followed by computing the total energy 
%change to decide whether to proceed with the update. 
To follow the temperature evolution of FE and AFD modes and lattice parameters, 
we used simulated annealing technique. The simulation supercell 
$20\times20\times20$ of BiAlO$_3$ unit cells was used with   periodic 
boundary conditions applied in all three directions to simulate bulk sample. 
Annealing started at  1500~K and  proceeded in steps of 
10~K until the temperature reaches 10~K.
We used 10 million MC steps for temperatures in the range 1350 to 1080 K in the 
vicinity of phase transition and 0.1 million MC steps for rest of the temperature steps. 
The larger MC steps in the vicinity of phase transition was used to increase the 
simulation time to capture the competing instabilities more accurately. 
First 60\% of these steps were used for equilibration, and the rest for computing averages.

%%%%%%%%%%%%%%%%%%%%%%%%%%%%%%%%%%% table
\begin{table*}
\caption{Interaction parameters of the effective Hamiltonian of BiAlO$_3$ derived 
	from first-principles density functional theory calculations.
	All parameters are listed in atomic units in the notations of 
	Refs. \cite{zhong-95, vanderbilt-98}. The equilibrium lattice parameter 
	for cubic phase is obtained as 7.03225 (a.u.), and the normalized 
	distortions for $\Gamma + R_{25}$ mode are: \(\xi_{\text{Bi}}\) = -0.591508, 
	\(\xi_{\text{Al}}\) = -0.032125, \(\xi_{\text{O}_1}\) = -0.015275, 
	\(\xi_{\text{O}_2}\) = 0.774787, \(\xi_{\text{O}_3}\) = -0.220361.}
\begin{ruledtabular}
\begin{tabular}{lcrcrcr}
 Interaction &  & & Parameters' values &  & &  \\ \hline
 FE onsite&$\kappa_2$&$0.019443$&$\alpha$&$0.040325$&$\gamma$&$-0.019488$ \\ \\
 
  &$j_1$&$-0.009872$&$j_2$&$0.024738$ \\
 FE intersite&$j_3$&$-0.000244$&$j_4$&$-0.002173$&$j_5$&$-0.001751$ \\
  &$j_6$&$0.000731$&$j_7$&$0.000001$ \\ \\

 Elastic&$B_{11}$&$4.379691$&$B_{12}$&$1.686690$&$B_{44}$&$1.782375$\\ \\

 FE-strain coupling&$B_{1xx}$&$-0.55062$&$B_{1yy}$&$-0.166299$&$B_{4yz}$&$-0.109343$\\ \\

 FE dipole&$Z^*$&$8.164117$&$\epsilon_\infty$&$8.061037$\\ \\

 AFD onsite&$\tilde{\kappa}_2$&$0.006134$&$\tilde{\alpha}$&$0.043132$&$\tilde{\gamma}$&$-0.026356$\\ \\

  &$\tilde{j}_1$&$0.007741$&$\tilde{j}_2$&$-0.002177$ \\ 
 AFD intersite&$\tilde{j}_3$&$-0.000295$&$\tilde{j}_4$&$-0.010826$&$\tilde{j}_5$&$0.008305$ \\
  &$\tilde{j}_6$&$0.000840$&$\tilde{j}_7$&$-0.003985$\\ \\

 AFD-strain coupling&$\tilde{B}_{1yyx},\tilde{B}_{2yyx}$&$-0.002980$&$\tilde{B}_{3yyx}$&$-0.066656$&$\tilde{B}_{4yzx}$&$0.013998$ \\ \\

 FE-AFD coupling&$G_{xxxx}$&$0.057239$&$G_{xxyy}$&$0.089787$&$G_{xyxy}$&$-0.169174$ \\
\end{tabular}
\end{ruledtabular}
\label{tab_param}
\end{table*}

%%%%%%%%%%%%%%%%%%%% FIGURE 2 
\begin{figure*}
\begin{center}
  \includegraphics[width = 0.6\textwidth]{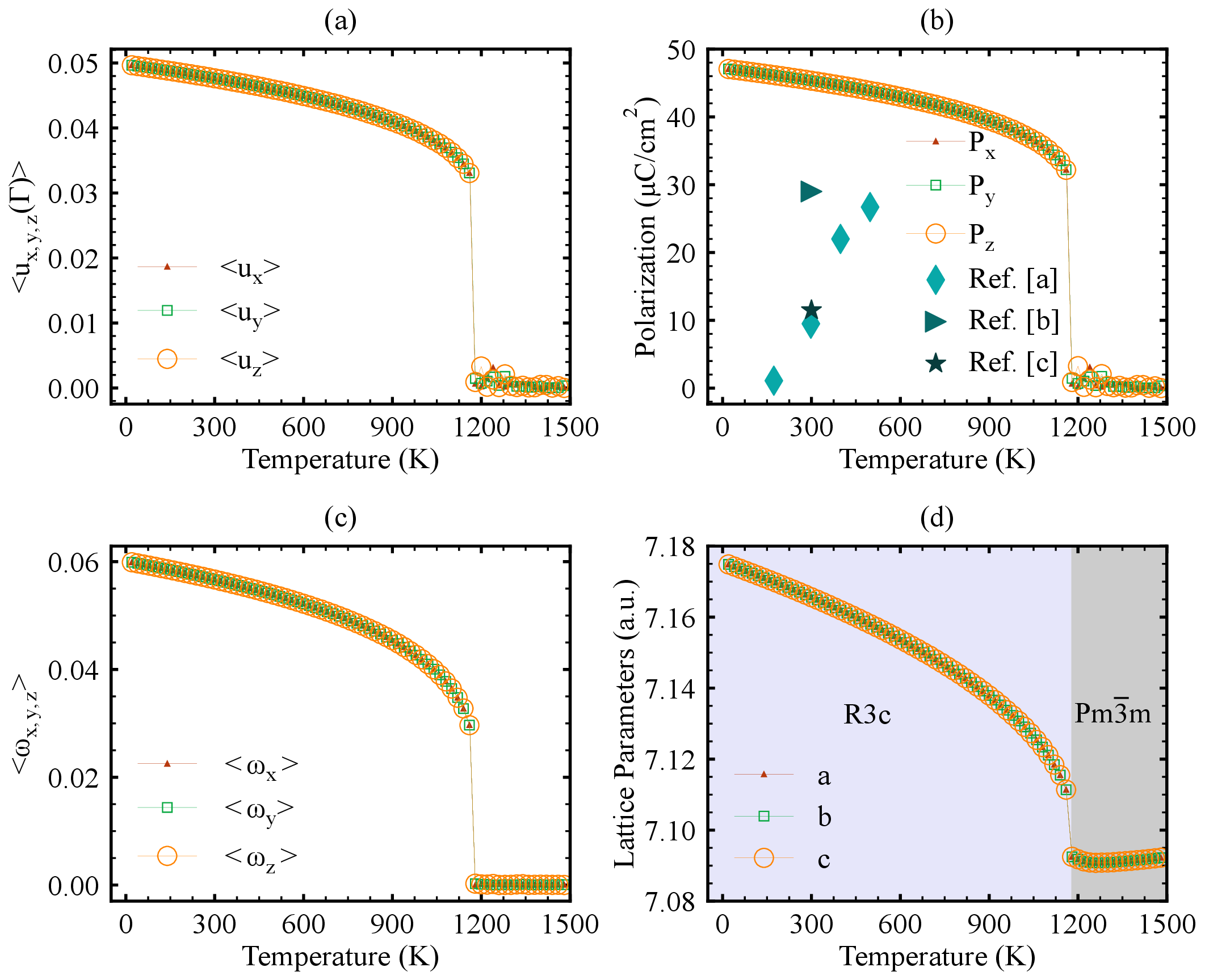}
	\caption{Temperature evolution of FE order parameter (panel (a)), 
	polarization (panel (b)), AFD order parameter (panel (c)) 
	and lattice parameter (panel (d)) obtained from the MC 
	simulation. Refs. [a], [b], and [c] corresponds to the 
	experiments \cite{zylberberg-07}, \cite{son-08}, and \cite{mangalam-08}, 
	respectively.}
  \label{fig_op}
\end{center}
\end{figure*}  

\section{Finite-temperature properties}

Figure \ref{fig_op} shows the temperature evolution of FE and AFD order 
parameters obtained from MC simulations. 
As discernible from Fig. \ref{fig_op}(a), the mean local-mode vectors 
along all three Cartesian directions ($u_x, u_y, u_z$) are close to zero at 
higher temperatures; however, as the system is annealed down to around 1200 K 
they start increasing significantly, leading to a structural phase transition. 
Consistent with the experimental observations for 
bulk \cite{zylberberg-07,mangalam-08} and film \cite{son-08}, 
our simulations predict a single phase FE transition (at 1160 K) from 
the centrosymmetric cubic ($Pm\bar3m$) to a non-centrosymmetric 
rhombohedral ($R3c$) phase.  
The local mode data are converted to 
polarization and plotted in panel (b) of the figure.  We observe a 
spontaneous polarization of $\sim$81 $\mu C/{\rm cm}^2$ along [111] 
direction at 10 K. This is consistent with the DFT values, $\sim$ 73 $\mu C/{\rm cm}^2$ 
from our calculation and $\sim$ 75.6 $\mu C/{\rm cm}^2$ from Ref. \cite{baettig-05}, 
at zero Kelvin. The slightly higher value from effective Hamiltonian 
could be attributed to the approximations used in constructing 
Eq.( \ref{h_eff}), as only dominant distortions are considered.
There is, however, a large variation in the polarization values reported in 
experiments \cite{mangalam-08, zylberberg-07, son-08}.
This trend in experimental and theoretical values of polarization 
suggests the relooking of polarization in BiAlO$_3$.
As discussed earlier, Curie temperature obtained from our simulations 
is 1160 K, which is greater than the experimentally observed value 
of $793$ K \cite{zylberberg-07}. This discrepancy could be attributed 
to an overestimation of the binding energy of the structure used in the 
parameterization of effective Hamiltonian.
As can be observed from panel (c), the AFD mode mimics the same trend as FE, 
with AFD order parameter ($<\omega_{x,y,z}>$) oriented 
along [111] pseudocubic direction (with $R\bar3c$ symmetry) under the 
phase transition. 
In order to assess the impact of interaction between FE and AFD modes, 
we repeated annealing simulations by switching off the FE-AFD interaction.
From our simulations, we observed both FE and AFD modes to be competing instabilities.
Panel (d) shows the variation of lattice parameters as a function of 
temperature. Consistent with FE and AFD order parameters, it exhibits 
a first-order-like phase transition with almost no change across 
the Pm$\Bar{3}$m phase and a smooth increase below the $T_{\rm C}$.

%%%%%%%%%%%%%%%%%%%%%%%%% FIGURE 3 
\begin{figure*}
\begin{center}
	\includegraphics[width = 0.7\textwidth]{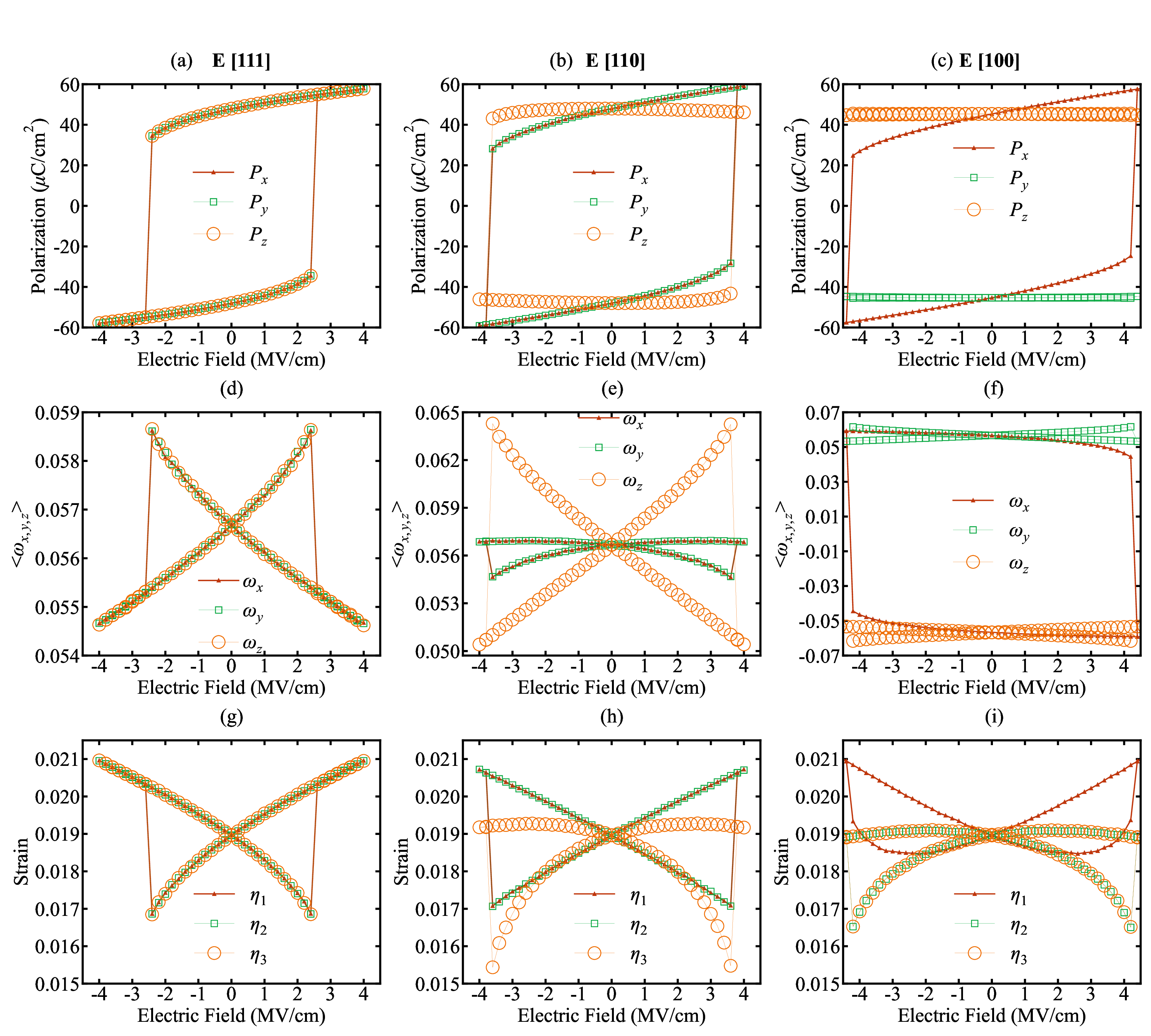}
	\caption{Electric field evolution of polarization (panels (a) to (c)), AFD 
	order parameters (panels (d) to (f)), and strain order parameters 
	(panels (g) to (i)) for the fields along [111], [110], and [100] 
	directions at room temperature. The AFD order parameters are in 
	terms of the units of the cubic lattice parameter of BiAlO$_3$. }
  \label{fig_hys}
\end{center}
\end{figure*}

Next, we examine the effect of applied electric field on the properties
of BiAlO$_3$. To assess this, an interaction term, 
${-\sum_{i} \mathbf p_i\cdot {\mathbf E}}$, was added to the effective 
Hamiltonian, where $\mathbf p_i$ is represented in terms of the Born 
effective charge and the local mode.
To understand the complete trend, we performed three distinct sets of MC 
simulations where we applied a {\em dc} electric field along [111], [110], 
and [100] pseudocubic directions. Throughout these simulations, the electric 
field is gradually introduced, subsequently withdrawn, followed by reversing 
its direction, and then finally removed again. 
Fig. \ref{fig_hys} shows the simulation data on electric field evolution 
of FE, AFD, and strain order parameters at 300 K. 
We observe a single hysteresis loop for all the components of polarization 
for the field along [111] direction (panel (a)). This confirms a ferroelectric 
$R3c$ phase of BiAlO$_3$ at room temperature.
For the field along [110] direction, the material still exhibits ferroelectric 
hysteresis, however, as can be expected, the $z$-component of polarization 
remains unaffected by the field and stays almost same as $P_{\rm S}$.  
Following the trend of [110], for a field along [100] direction, we observed 
hysteresis for $x$-component of polarization, whereas polarization vectors along
$y-$ and $z$-directions remain unaffected by the field. 
In terms of the magnitude of $P_{\rm Sat}$, there is a slight 
change in the values for the fields along different directions. We obtained
$P_{\rm Sat}$ of 57.8, 59.2, and 60.2 $\mu$C/cm$^2$ for the fields 
along [111], [110], and [100] directions, respectively. These are comparable 
to technologically relevant lead-based ferroelectrics \cite{JUN-KI_2007, mani-13}.   
However, as can be expected, we observe a significant variation in 
the coercive fields ($E_{\rm C}$) for different field directions; we get 
$E_{\rm C}$ as 2.4, 3.6, and 3.8 MV/cm for the aforementioned
sequence of field directions. The differences in $E_{\rm C}$ along
different field directions highlight the anisotropic nature of the 
FE phase of BiAlO$_3$.

Panels (d), (e), and (f) show the evolution of AFD order parameters under 
the influence of applied electric fields in different directions. We observe 
a butterfly loop like behavior with AFD vector pointing parallel to FE 
vector for the field along [111] direction. For the field along [110], we get
a similar behavior, however, with two key differences. First, consistent 
with the larger $E_{\rm C}$ for FE mode, the transition field is larger. 
Second, as can be expected due to confinement of $R_{25}$ distortions in 
xy-plane, the z-component of AFD vector decouples. Interestingly, 
for the field along [100], we observe a similar decoupling as [110], however, 
with a change in the phase of $<\omega_z>$. The field evolution of strain 
tensor is shown in the panels (g), (h), and (i).
As discernible from the figures, strains appear to preferentially align 
in the field directions and exhibit a butterfly like feature; however, 
like the case of FE and AFD order parameters, with a decoupling of 
$\eta_3$ for [110] and [100] fields. Notably, a high saturation 
strain of 0.021 and a significant remanent strain ranging from 0.018 
to 0.019 are observed. Such a high value of remanent strain could enable 
BiAlO$_3$ as a promising candidate for shape memory applications \cite{Cheikh_2016}. 
Additionally, a significant strain change of 0.004 is observed in 
the vicinity of critical field when polarization switches the direction.  

%%%%%%%%%%%%%%%%%%%%%%%%% FIGURE 4 
\begin{figure*}
\begin{center}
  \includegraphics[width = 0.7\textwidth]{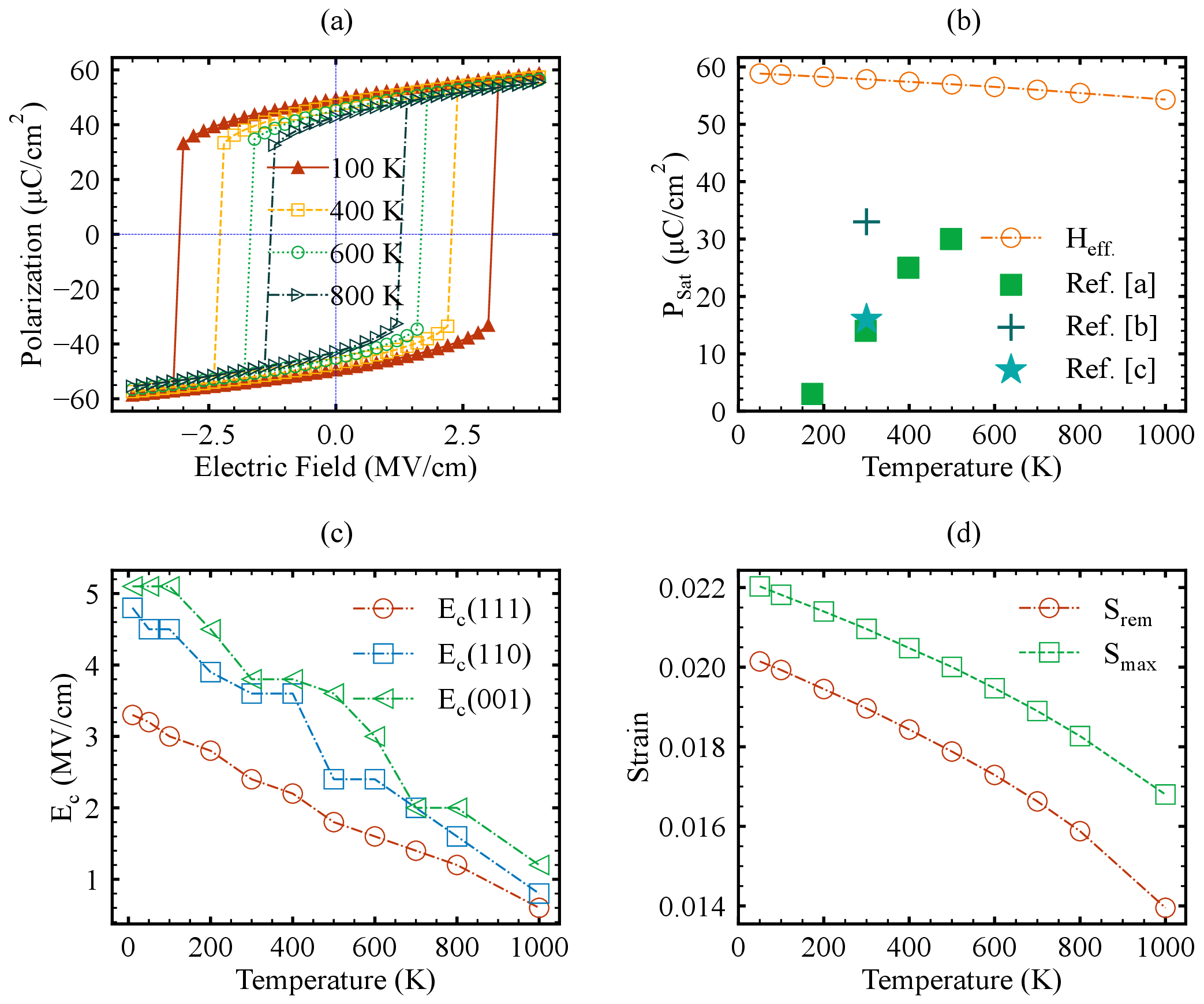}
	\caption{(a) Temperature dependent P-E hysteresis for fields 
	along [111] direction. (b), (c) Variation of saturation polarization 
	and coercive field as function of temperature. 
	(d) Maximum and remanent strains as a function of temperature. Refs. [a],
	[b], and [c] correspond to the experiments \cite{zylberberg-07}, \cite{son-08}, 
	and \cite{mangalam-08}, respectively.}
    \label{fig_hys_t}
\end{center}
\end{figure*}

Next, we probe the response of applied electric field along [111] direction 
for a wide range of temperatures. Figure \ref{fig_hys_t} (a) shows the 
temperature dependent hysteresis for temperatures ranging from 100 to 800 K.  
Consistent with other prominent ferroelectric 
materials \cite{Wenhui_2014,HfO2_2022,Patel_2015,PMN_PTO_2017,PMNO_PTO_2005,PINO_PTO_2017,PMN_PT_2022},
and as can be expected, the area enclosed by the hystereses reduce
when the operating temperature increases, and as a result, it leads to
decrease in the $P_{\rm S}$, $P_{\rm Sat}$ and $E_{\rm C}$ 
values with increasing temperature. 
The variation of $P_{\rm Sat}$ as function of operating 
temperature is shown in panel (b) for a wide range of temperatures. 
The observed decreasing trend of our results exhibit an opposite behavior
to the trend of the only available experimental study \cite{zylberberg-07}. 
The reason for this 
opposite behavior of experimental data could be attributed to the fact 
that $P_{\rm Sat}$ values reported in Ref. \cite{zylberberg-07}
are not fully saturated due to limitations in applying the larger
fields. The experimentally reported values of $P_{\rm Sat}$ at room temperature 
are 33 \cite{son-08}, 16 \cite{mangalam-08} and 14 $\mu$C/cm$^2$ \cite{zylberberg-07}.  
Our computed value, 58$\mu$C/cm$^2$, at 300 K is higher than the experiments. 
The reason for this could be ascribed to the defects and grain boundaries 
present in the experimental sample which can cause the leakage current.

%%%%%%%%%%%%%%%%%%%%%%%%% FIGURE 5 
\begin{figure*}
\begin{center}
  \includegraphics[width = 0.8\textwidth]{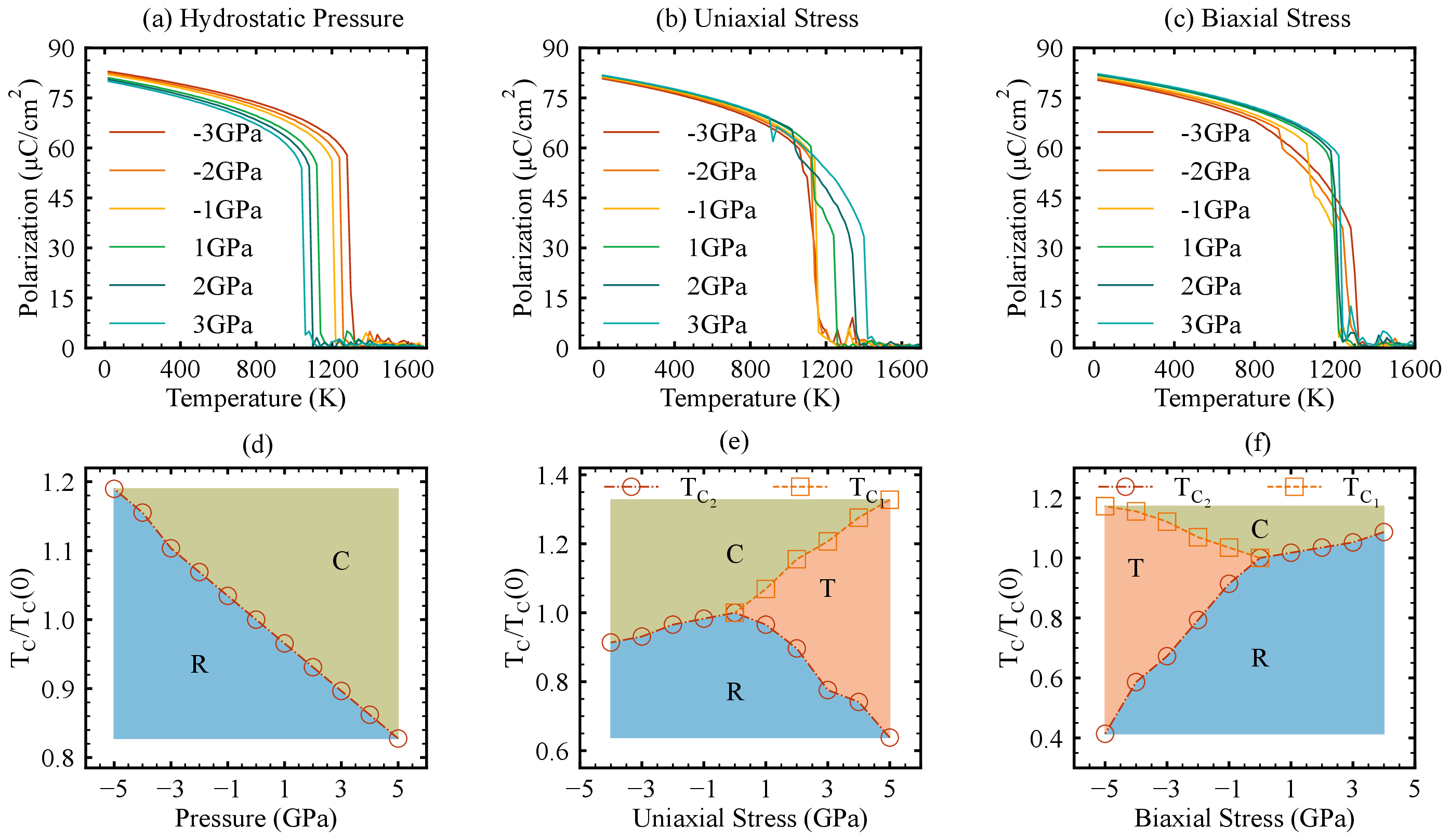}
  \caption{Evolution of polarization as a function of temperature at 
	different hydrostatic pressure (panel (a)), uniaxial (panel (b)) 
	and biaxial (panel (c)) stress. Evolution of Curie temperature 
	as function of hydrostatic pressure (panel (d)), uniaxial (panel (e)) 
	and biaxial (panel (f)) stress.}
  \label{fig_pres}
\end{center}
\end{figure*}

The temperature evolution of E$_{\rm C}$ is shown in panel (c) of Fig. \ref{fig_hys_t}. 
As discernible from the figure, for all three directions of the applied 
field, E$_{\rm C}$ decreases with increasing temperature. This is consistent 
with the trends observed in other ferroelectric materials \cite{ducharne-00, mani_pzo_2015}.
The observed temperature dependence could be attributed to a combination of 
factors like polarization anisotropy and temperature-dependent domain wall dynamics.  
The observed trend from our simulations is consistent with other promising 
ferroelectrics \cite{PMNO_PTO_2005,PINO_PTO_2017,PMN_PT_2022,kashikar_coexistence_2024}.
However, our simulations tend to overestimate the coercive fields as compared 
to experiments \cite{zylberberg-07, mangalam-08,son-08}. As mentioned earlier, 
the reason could be attributed to the factors like absence of defects within 
the simulated structure. It is to be noted that, defect free samples undergo 
homogeneous phase transitions without forming any domain, and hence expected 
to exhibit a high coercive field. 
Panel (d) demonstrates the temperature response of remanent (S$_{\rm rem}$) and 
maximum (S$_{\rm max}$) strains for [111] field direction.  
As discernible from the figure, both S$_{\rm rem}$ and S$_{\rm max}$ 
decrease with increasing temperature. This is consistent with the experimental 
observations for other prominent ferroelectrics \cite{Mohan_2016,Wenhui_2014,Narit_2020}.

Next, we examine the effects of hydrostatic pressure and stress on $P_{\rm Sat}$ 
and $T_{\rm C}$. As discernible from Fig. \ref{fig_pres}(a), we observe
a single phase transition for the entire range of applied hydrostatic pressure. 
The $T_{\rm C}$ is found to decrease(increase) with an application of 
positive(negative) pressure with a rate of -45 K/GPa (panel (d)). While 
this observed trend is consistent with ferroelectrics \cite{McCash-pto}, 
it is of the opposite nature to that observed in the antiferroelectrics \cite{mani_pzo_2015}. 
The observed change in $T_{\rm C}$ in response to pressure provides a scope for tuning 
it depending on the need for device applications. We also notice a slight 
decrease in the $P_{\rm S}$ values, with a rate of 0.5 $\mu$C/cm$^2$/GPa, 
from negative to positive pressure.

Panels (b) and (c) show the stress response of $P_{\rm S}$ along [111] 
direction. For this, we simulated the uniaxial and  biaxial stresses in the range 
$-5.0$ GPa to $5.0$ GPa in a step of $1.0$ GPa. The effect of uniaxial 
stress was simulated by applying a single component of the stress 
tensor ($\sigma$), whereas for biaxial stress, we fixed 
$\sigma_1 = \sigma_2 = \sigma$ and $\sigma_6 = 0$, while allowing all 
other components to relax.
We find, under the tensile uniaxial stress, material exhibits 
a sequence of two phase transitions (panel (b)); initially from cubic to 
tetragonal and then from tetragonal to rhombohedral. 
On contrary, for the compressive uniaxial stress, the material is observed 
to undergo a single phase transition from cubic to rhombohedral. 
Examining the effect of biaxial stress, we observe an opposite trend; 
a single phase transition for tensile and two phase transitions for 
compressive stress (panel (c)). 
As discernible from panel (e) of the figure, uniaxial tensile stress 
has more significant impact on transition temperature as compared to 
the compressive counterpart. The transition temperature for cubic to 
tetragonal ($T_{\rm C_1}$) increases rapidly with increasing uniaxial 
tensile stress. Whereas, the transition temperature for tetragonal 
to rhombohedral ($T_{\rm C_2}$) decreases with tensile stress. 
Transition temperatures $T_{\rm C_1}$, $T_{\rm C_2}$ exhibit an 
opposite trend under biaxial stress (panel (f)) as compared to 
uniaxial counterpart.

%%%%%%%%%%%%%%%%%%%%%%%%%%%%%%%%%%%%%%%%%%%%%%%%%%%%%%%%%%%%%%%%%%%%%
%%%%%%%   Section: Conclusion                                  %%%%%%
%%%%%%%%%%%%%%%%%%%%%%%%%%%%%%%%%%%%%%%%%%%%%%%%%%%%%%%%%%%%%%%%%%%%%
\section{Conclusion}

We have developed a first-principles based atomistic model
to systematically simulate the properties of lead-free perovskite BiAlO$_3$ 
at finite temperatures. The model was then employed to probe the 
characteristics of the phase transition and ferroelectric properties in 
bulk BiAlO$_3$. Consistent with the experimental observations, our simulations
predict that BiAlO$_3$ undergoes a paraelectric to ferroelectric phase 
transition, with $T_{\rm C}$ of 1160 K. 
The observed ferroelectric phase is of rhombohedral nature, associated 
with $R3c$ space group and spontaneous polarization of 81 $\mu$C/cm$^2$ 
along [111] direction. The AFD mode is found to obey the same trend of 
phase transition as FE with AFD order parameter oriented along the FE vector.
Our simulations on electric-field-dependent properties show a perfect 
hysteresis for polarization and butterfly-like loops for AFD and strain as
function of field along [111] direction. Our simulations
predict a high value of saturation and remanent strains, which could be 
suitable for shape memory applications. From the temperature dependent 
hysteresis simulations, we observed a decrease in the values of saturation 
and remanent polarizations and coercive field with increasing temperature. 
Both saturation and remanent strains are observed to decrease with 
temperature. From the pressure-dependent study, we observed a decrease 
in the spontaneous polarization and Curie temperature from negative 
to positive hydrostatic pressure. Application of uniaxial and biaxial 
stresses are found to introduce new phases of FE in the material.

%%%%%%%%%%%%%%%%%%%%%%%%%%%%%%%%%%%%%%%%%%%%%
\begin{acknowledgments}

BKM acknowledges the funding support from SERB, DST (CRG/2022/000178). 
The calculations were  performed using the High Performance Computing 
cluster Tejas at the Indian Institute of Technology Delhi and 
PARAM Rudra, at IUAC, New Delhi.

\end{acknowledgments}

%%%%%%%%%%%%%%%%%%%%
\bibliographystyle{apsrev4-2}
\bibliography{BAO}

\end{document}